\documentclass[twocolumn,pra,showpacs,floatfix]{revtex4}

\usepackage{amsmath}
\usepackage{amsfonts}
\usepackage{amsthm}
\usepackage{graphicx}
\usepackage{amssymb}
\newcommand{\ket}[1]{\ensuremath{|#1\rangle}}
\newcommand{\bra}[1]{\ensuremath{\langle #1|}} 

\newcommand{\axis}{\ensuremath\mathbf{\hat{n}}}
\newcommand{\nuc}[2]{\mbox{${}^{#1}\rm #2$}}
\newcommand{\units}[2]{\mbox{$#1\,\text{#2}$}}

\begin{document}
\title{Practical Implementations of Twirl Operations}
\author{M.~S. Anwar}
\email{muhammad.anwar@physics.ox.ac.uk} \affiliation{Centre for
Quantum Computation, Clarendon Laboratory, University of Oxford,
Parks Road, OX1 3PU, United Kingdom}

\author{L.~Xiao}
\affiliation{Centre for Quantum Computation, Clarendon Laboratory,
University of Oxford, Parks Road, OX1 3PU, United Kingdom}

\author{A.~J. Short}
\affiliation{Centre for Quantum Computation, Clarendon Laboratory,
University of Oxford, Parks Road, OX1 3PU, United Kingdom}

\author{J.~A. Jones}
\email{jonathan.jones@qubit.org} \affiliation{Centre for Quantum
Computation, Clarendon Laboratory, University of Oxford, Parks
Road, OX1 3PU, United Kingdom}

\author{D.~Blazina}
\affiliation{Department of Chemistry, University of York,
Heslington, York, YO10 5DD, United Kingdom}

\author{S.~B. Duckett}
\email{sbd3@york.ac.uk} \affiliation{Department of Chemistry,
University of York, Heslington, York, YO10 5DD, United Kingdom}

\author{H.~A.~Carteret}
\email{cartereh@iro.umontreal.ca} \affiliation{LITQ, Departement
d'Informatique et Recherche Op\'erationelle, Pavillon
Andr\'e-Aisenstadt, Universit\'e de Montr\'eal, Montr\'eal,
Qu\'ebec H3C 3J7, Canada}

\date{\today}

\begin{abstract}
Twirl operations, which convert impure singlet states into Werner
states, play an important role in many schemes for entanglement
purification.  In this paper we describe strategies for
implementing twirl operations, with an emphasis on methods
suitable for ensemble quantum information processors such as
nuclear magnetic resonance (NMR) quantum computers.  We implement
our twirl operation on a general two-spin mixed state using liquid
state NMR techniques, demonstrating that we can obtain the singlet
Werner state with high fidelity.
\end{abstract} \maketitle

\section{Introduction}
The twirl operation was introduced by Bennett \textit{et al.}
\cite{bennett96,bennett96b} in the context of entanglement
purification of mixed states. The first step in many of these
protocols is the twirl operation which converts an arbitrary mixed
state of a two qubit system, $\rho$, into a Werner singlet state
$\rho_W$ \cite{werner89}, an incoherent mixture of $\varepsilon$
parts of a maximally entangled singlet state,
$\ket{\Psi^-}=(\ket{01}-\ket{10})/\sqrt{2}$, and $(1-\varepsilon)$
parts of the maximally mixed state,
\begin{equation}\label{werner}
\rho_{W}(\varepsilon)=\varepsilon\ket{\Psi^-}\bra{\Psi^-}+(1-\varepsilon)\frac{\openone}{4}.
\end{equation}
The key idea behind the twirl is that the singlet state is invariant
under any bilateral unitary transformation of the two qubits (that
is, any operation where identical local unitaries are applied to the
two qubits), whereas any other state will be affected.  A randomly
chosen bilateral rotation should serve to average any other state,
converting an arbitrary mixed state to a Werner singlet.  (Note that
the twirl sequence does not \textit{create} new singlet, rather it
\textit{preserves} the singlet, while averaging out all other
terms.) This provides a good ``in principle'' definition of the
twirl operation, but in practice it is important to consider what a
random bilateral rotation really is, and how it can actually be
implemented.

Early work in this field concentrated on reducing the infinite set
of randomly chosen rotations down to a small finite set. Originally
it was suggested \cite{bennett96} that the twirl could be achieved
using a set of four bilateral rotations, although it was later shown
that a set of twelve bilateral rotations are both necessary and
sufficient \cite{bennett96b}.  This approach is sensible for
conventional quantum information processors, but with ensemble
devices, such as nuclear magnetic resonance (NMR) quantum computers,
it is useful to use a different approach.  In such devices it is in
fact easier to perform rotations by angles that vary continuously
over the ensemble than by angles from some small carefully chosen
set. In NMR devices, for example, the application of a strong
magnetic field gradient \cite{keeler} performs a rotation around the
$z$ axis with a rotation angle that depends strongly on the position
of a given molecule within the spatial ensemble; integrating over
the position within the sample is then equivalent to applying a
random rotation.  In the language of NMR this is often referred to
as a \textit{crush gradient} pulse. For simplicity we will sometimes
describe a crush gradient as a random rotation around the $z$-axis,
as the effect on the averaged density matrix is the same, but when
considering experimental implementations it is necessary to be more
precise.

The use of field gradients forms the basis of spatial averaging
methods, used in NMR quantum computing to prepare pseudo-pure
states \cite{cory96, jones01a}. Field gradients have also been
used to average out error terms \cite{jones98b}, to project qubits
into the Zeeman basis \cite{nielsen98}, and to simplify density
matrices prior to partial state tomography \cite{anwar03}. The
alternative procedure, in which one rotation chosen from a small
set is applied, requires several experiments to be performed, and
in the context of NMR quantum computing is usually known as
temporal averaging.

In subsequent sections we explore different methods for implementing
the twirl, concentrating on those which are best suited to ensemble
techniques.  We begin by noting that any set of bilateral rotations
which performs a twirl must correspond to a set of single qubit
rotations which, when considered from the viewpoint of a single
qubit subsystem, averages any state of a single qubit to the
maximally mixed state. As we shall see, however, averaging a single
qubit is a necessary but not a sufficient condition for a set of
rotations to act as a twirl.

\section{Averaging a single qubit}
Averaging a single qubit gives some useful insight into the problem
of twirling a two qubit state. A single qubit state can be
represented by a ray on the Bloch sphere and an arbitrary rotation
will move this ray over the surface of the sphere. Each rotation
$U(\xi ,\mathbf{\hat{n}})$ is defined by a rotation angle $\xi$
about an axis $\axis$ parameterized by a tilt angle $\theta$ and an
azimuthal angle $\phi$. An arbitrary single qubit state is
completely defined by three real parameters, which are
conventionally taken as the expectation values of the single spin
Pauli operators $\{\sigma_x,\sigma_y,\sigma_z\}$ \cite{nielsen00}.
For scrambling an arbitrary state to the maximally mixed state, we
must choose a convenient set of operations, which scrambles each of
these Pauli operators individually. A tempting, but incorrect,
approach is to apply random rotations about random axes, that is to
take $\xi$ as being uniformly distributed between $0$ and $2\pi$,
while the rotation axes defined by $\theta$ and $\phi$ are uniformly
distributed over the sphere; we call this set of rotations
$\mathcal{R}$.

Brute force integration shows that the continuous set of random
rotations around random axes, $\mathcal{R}$, does not completely
randomize the state of a single qubit, and so \textit{cannot}
constitute the basis for a twirl operation; instead it reduces the
Bloch vector to one third of its original length. In retrospect the
reason for this behaviour is clear: random rotations around axes
perpendicular to the original state will completely average it,
while rotations around axes parallel or anti-parallel to the state
will leave it unaffected. Thus the overall effect of $\mathcal{R}$
must be to scale down the state, rather than to average it
completely.  The significance of the scaling by one third is
discussed below.

A better definition of a random rotation is provided by
considering the description of a rotation by means of its Euler
angles.  While many different conventions for describing Euler
angles exist, the essential feature is that any rotation can be
decomposed as a sequence of three rotations around two axes.  For
example any rotation can be achieved by the sequence of rotations
\begin{equation}
R_z(\phi)\,R_y(\theta)\,R_z(\xi)
\end{equation}
where rotations are applied from left to right.  A random
distribution of Euler rotations can be achieved by taking $\xi$ as
uniformly distributed between $0$ and $2\pi$, with the rotation
axes defined by $\theta$ and $\phi$ uniformly distributed over the
sphere; we call this set of rotations $\mathcal{E}$.  It might
seem that $\mathcal{E}$ is the same as $\mathcal{R}$, but they are
in fact quite different: in particular $\mathcal{E}$ completely
averages a single qubit, while $\mathcal{R}$ does not.  This is
easily seen by noting that the first rotation will average
$\sigma_x$ and $\sigma_y$, while the second and third will average
$\sigma_z$.  As any state of a single qubit can be written as a
linear combination of these basic matrices and the maximally mixed
state, the first two operations will average \textit{any} state.

It is clear from the above that a random Euler rotation will
average any single qubit, but this process is perhaps excessive,
as two rotations suffice.  (We will see below that this is not
true when considering a true twirl applied to two qubits.) The
sequence of rotations
\begin{equation}
R_z(\phi)\,R_y(\theta), \label{rotzy}
\end{equation}
where both $\theta$ and $\phi$ are now taken as uniformly
distributed between $0$ and $2\pi$, will average any single qubit;
as before this is most simply seen by considering the result for
$\sigma_x$, $\sigma_y$ and $\sigma_z$.  Indeed this process can be
simplified still further: as described by Bennett \textit{et al.}
\cite{bennett96} is is possible to average a single qubit by
randomly selecting from the four operations $\{\openone, \sigma_x,
\sigma_y, \sigma_z\}$. Since $\sigma_\alpha$ is equivalent to a
$180^\circ_\alpha$ rotation, and (neglecting global phases)
$\sigma_x=\sigma_y\sigma_z$ it is clear that this operation is
equivalent to applying either $0$ or $180^\circ_z$ at random, and
then applying $0$ or $180^\circ_y$ at random; a similar result has
been described by Hayden et al \cite{hayden}. Thus instead of
choosing the two angles in equation \ref{rotzy} at random from
uniform distributions, we can choose at random from two particular
values.

This process of replacing a continuous rotation by a small number
of discrete values is an example of a quite general procedure. For
rank $1$ and rank $2$ tensors, continuous rotations about a fixed
axis, can in fact, always be replaced by discrete rotations: we
can ``hop'' between the angles $2\pi k/p$, with $p\ge 3$ and
$k=1,2,\ldots ,p$, and the overall effect is the same as rotating
continuously about the same axis \cite{schmidt-rohr}.  It is
\textit{sometimes} possible to use $p=2$, but this cannot be
relied on in general.

The rotation through $2\pi k/p$ is the member $C_p^k$ of the
cyclic group $\mathbb{C}_p$ \cite{jones}; in this way, the
continuous rotations can be replaced by discrete group operations
and the continuous integrals over the rotation angle $\xi$, can be
substituted by \textit{group theoretical averages}, such as
\begin{equation}
\int_{\xi}U(\xi,\axis)\rho U^{\dagger}(\xi,\axis)\,d\xi\equiv
\sum_{k=0}^{p-1} U(\frac{2\pi k}{p},\axis)\rho
U^{\dagger}(\frac{2\pi k}{p},\axis).
\end{equation}
The simplest discrete group that can generally be used for
averaging is the cyclic $\mathbb{C}_3$ group, which is employed by
NMR spectroscopists in the form of magic angle hopping
\cite{schmidt-rohr,bax83}, a discrete version of its continuous
counterpart, magic angle spinning \cite{schmidt-rohr}.  The group
theoretical framework has also been used in the context of
bang-bang decoupling protocols \cite{viola99,viola04}, which seek
to average out system-environment interactions by applying fast,
discrete and periodic control impulses to \textit{only} the
system, with the control propagators faithfully representing the
members of a discrete cyclic group $\mathbb{C}_n$.

This approach hints at another method for performing complete
averaging of a single qubit.  We have already noted that random
rotations around a random axis do not completely average a single
qubit, but simply rescale its length by one third.  This process is,
as usual, equivalent to applying a rotation through an angle of
$0^\circ$, $120^\circ$ or $240^\circ$ around a random axis. Clearly
a $0^\circ$ rotation can have no effect, while the effect of the
$120^\circ$ and $240^\circ$ rotations must be the same. From this it
can be deduced that a $120^\circ$ rotation around a random axis will
completely average a single qubit, and this is indeed the case.

Finally we turn to the issue of practical experimental
implementations.  For a conventional quantum information processor
the obvious approach is to use the smallest discrete set of
operations; for the case of averaging a single qubit this is the
set of four rotations described by Bennett \textit{et al.}  With
an ensemble processor, such as an NMR quantum computer, it is
better to use a procedure based on continuous rotations,
corresponding to spatial averaging.  Clearly the best approach is
to apply random rotations around two orthogonal axes, such as $z$
and $y$.  A strong magnetic field gradient, denoted \textbf{G},
will effect a random rotation around the $z$ axis, and rotations
around other axes can be achieved by combining gradients with
single qubit gates (radiofrequency pulses) which can be treated as
rotating the axis system.  Thus the sequence
\begin{equation}\label{one-qubit-crot-2axes}
\textbf{G}\,90_x\,\textbf{G}
\end{equation}
will completely average a single spin.  Note that this procedure
is not completely equivalent to equation \ref{rotzy}, as it should
strictly speaking be followed by a $90_{-x}$ pulse to rotate the
axes back to their original positions, but as the maximally mixed
state will not be affected by the rotation this final stage may be
safely omitted.

The discussion above assumes that the gradient pulses are
instantaneous, so that their sole effect is average the spin state
and evolution under any background Hamiltonian can be ignored.  In
fact, gradient pulses take a finite time, but as the gradient and
background Hamiltonians commute they can be treated as an
instantaneous pulse followed by a finite period of evolution under
the background Hamiltonian.  For a single NMR spin this is not a
problem, as the maximally mixed state does not evolve under the
background Hamiltonian; for a two qubit system, however, it is
necessary to consider this problem more carefully.

\section{Twirling two qubits}
We now turn to the problem of implementing a full twirl on two
qubits.  This is clearly related to the problem of averaging a
single qubit, but is more complicated.  In particular, any twirl
procedure will average a single qubit, but not every averaging
procedure can be converted to a twirl.  For example, the sequence
of four operations $\{\openone, \sigma_x, \sigma_y, \sigma_z\}$
suggested by Bennett \textit{et al.} \cite{bennett96} does average
a single qubit, but does not effect a twirl; instead it is
necessary to use a set of 12 operations \cite{bennett96b}
corresponding to the rotational symmetry elements of the
tetrahedral group $T$ \cite{jones}.  The group theoretical
justification for using these rotations is discussed in
\cite{aravind97}.

Turning to the other averaging methods discussed above, only the
method based on random Euler rotations ($\mathcal{E}$) will
correctly implement a twirl; the other methods will only partially
average the state.  In particular the sequence of two random
rotations around orthogonal axes has already been studied
\cite{anwar03}, and shown to implement a partial twirl, in which
the density matrix is reduced to Bell diagonal form, but not to a
Werner state.

For an ensemble quantum information processor, it might seem that
the best strategy for performing a full twirl would be to perform
random Euler rotations.  Unfortunately this approach cannot be
implemented using gradients or similar methods, as these generate
rotations with a uniform distribution (between 0 and $2\pi$)
\textit{for each rotation angle}, while the Euler rotation method
requires that the angles $\theta$ and $\phi$ be uniformly
distributed \textit{over a sphere}.  Thus it might seem that twirl
operations cannot be easily implemented by ensemble methods.  It
is, however, possible to do this by using sequential random
rotations around three different axes.

This procedure is a development of the partial twirl described
above.  Consider again the sequence of rotations in equation
\ref{one-qubit-crot-2axes}. These two rotations constitute a
partial twirl \cite{anwar03}, and leave the density matrix Bell
diagonal with equal populations of $\ket{\Phi^+}$ and
$\ket{\Phi^-}$, where
$\ket{\Phi^\pm}=(\ket{00}\pm\ket{11})/\sqrt{2}$. The role of the
third rotation is to scramble these states with the sole remaining
undesired term, $\ket{\Psi^+}=(\ket{01}+\ket{10})/\sqrt{2}$,
leaving a Werner singlet state.  This can be achieved by rotating
about an axis at an angle $\psi=\arccos(1/\sqrt3)\approx
54.74^\circ$, commonly known in NMR studies as the ``magic angle''
\cite{schmidt-rohr,bax83}.  Our final twirl sequence is then
\begin{equation}\label{2-qubit-3-axes-Gzonly}
\mathbf{G}\,90_x\,\mathbf{G}\,54.74_x\,\mathbf{G}.
\end{equation}
As was the case for the partial twirl \cite{anwar03} it is
necessary to choose the lengths of the gradient pulses as
multiples of $1/\delta$, where $\delta$ is the difference between
the Larmor frequencies of the two spins, to refocus evolution
under the background Hamiltonian, which occurs at this difference
frequency.

If desired this twirl can be implemented using discrete steps by
replacing each random rotation with a three step averaging
procedure,
\begin{equation}
\{Z_3^m\}\,90_x\,\{Z_3^n\}\,54.74_x\,\{Z_3^p\},
\end{equation}
where $\{Z^a_b\}$ indicates a set of $z$ rotations with rotation
angles $2\pi\,a/b$, where $a=0,1,\ldots b-1$.  This discrete twirl
requires twenty-seven steps, but the number of steps can be
reduced to eighteen, using the sequences
\begin{equation}
\{Z_2^m\}\,90_x\,\{Z_3^n\}\,54.74^{\circ}_x\,\{Z_3^p\},
\end{equation}
or
\begin{equation}
\{Z_3^m\}\,90_x\,\{Z_2^n\}\,54.74_x\,\{Z_3^p\}.
\end{equation}
It is not possible to reduce the number of steps further while
using sequential rotations around three axes.

It is instructive to compare this discrete twirl with the twelve
step version of Bennett \textit{et al.} \cite{bennett96b}. Their
twelve rotations include the identity operation, $\pi$ rotations
about the principal $x$, $y$ and $z$ axes and rotations by $\pm
2\pi/3$ around the four body diagonals, each of which is at the
magic angle with the cardinal axes. (These are the twelve rotations
that leave a tetrahedron invariant.) Clearly our rotations are
closely related to theirs; the use of sequential rotations means
that the number of steps required is larger, but the sequential
approach may be simpler to implement in practice.

\section{An NMR Experiment}
Finally, we demonstrate the experimental implementation of our
ensemble twirl operation as described in equation
\ref{2-qubit-3-axes-Gzonly}. Our implementation is based on a
homonuclear system, comprising the two \nuc{1}{H} nuclei of
cytosine dissolved in $\text{D}_2\text{O}$; we use the product
operator description \cite{sorensen83} for the NMR states and
pulse sequences, and label the two spins as $I$ and $S$.  The
Hamiltonian of our system is then
\begin{equation}
\mathcal{H}/\hbar=2\pi\,\nu_I I_z + 2\pi\,\nu_S S_z + \pi
J\,2I_zS_z
\end{equation}
where $\nu_I$ and $\nu_S$ are the resonance offset frequencies of
spins $I$ and $S$, and $J$ is the $IS$ spin--spin coupling
constant (assuming weak coupling), all measured in Hertz.  NMR
experiments were performed on a Varian Unity INOVA
\units{600}{MHz} spectrometer with the spectrometer frequency
placed between the two cytosine resonances; for our system
$\nu_I=\units{457.9}{Hz}$, $\nu_S=\units{-457.9}{Hz}$ and
$J=\units{7.2}{Hz}$.

The experiment began with the preparation of a generic mixed state
of the two qubits containing a contribution from the singlet state
as well as a wide range of other terms. The preparation sequence
(which was applied to the thermal equilibrium state, $I_z+S_z$)
involves shaped pulses selectively exciting the spins and
separated by a delay $\tau$,
\begin{equation}\label{dirty-prepare}
\textbf{A}\equiv\,60I_y-\tau-30S_y.
\end{equation}
Shaped pulses were implemented using strongly modulated composite
pulses as described by Fortunato \textit{et al.} \cite{fortunato02}.
The deviation density matrix of the prepared state $\rho$ has a
singlet component proportional to
\begin{equation}\label{singlet-fraction-dirty}
\frac{\sqrt{3}}{8}\sin{(\pi J\tau)}\sin{(2\pi\nu_I\tau)}.
\end{equation}
Clearly the amount of singlet in $\rho$ can be easily controlled
through choice of the delay $\tau$.  Note that as we are using a
deviation density matrix description of a highly mixed state it is
possible for the singlet component to be negative.

We performed two sets of experiments on states drawn from this
family of states. The first set uses a fixed initial state and
demonstrates a stepwise progression through the twirl sequence, with
the resulting state becoming closer to a Werner state at each stage.
The second set uses a range of initial states, with different
singlet fractions, and shows that the twirl sequence works equally
well over this range.

\begin{figure*}
\begin{center}
\includegraphics[scale=1.0]{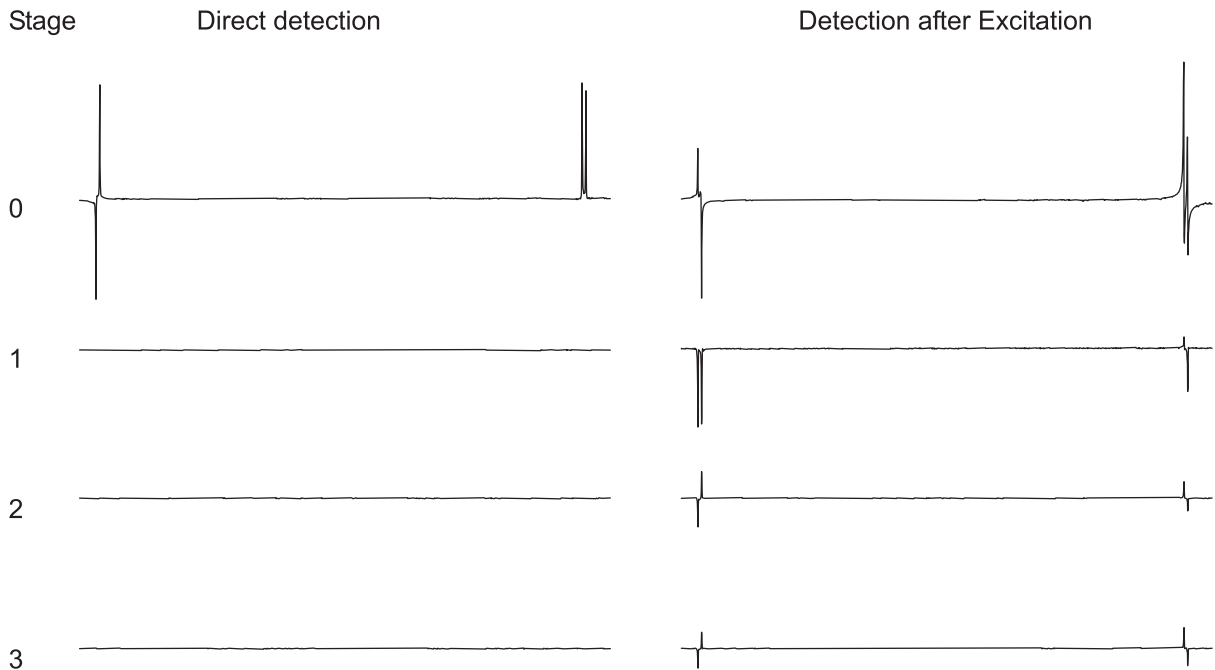}
\end{center}
\caption{Experimental spectra depicting an implementation of an
ensemble twirl.  Spectra were obtained using the pulse sequences
listed in table \ref{experiment1}.  The four rows correspond to
spectra acquired after 0, 1, 2 or 3 stages of the twirl sequence,
while the left and right columns correspond to direct signal
acquisition and acquisition after the selective excitation
sequence. The ideal result at the end of the full twirl (last row)
is no signal with direct detection (left column), and a pair of
antiphase doublets with equal and opposite intensities after a
selective excitation sequence (right column).}\label{graphics1}
\end{figure*}

\begin{table}
\caption{Pulse sequences for the first set of experiments, which
demonstrate the effects of the three stages of our twirl operation.
See main text for details.} \label{experiment1}
\begin{ruledtabular}
\begin{tabular}{cl}
Stage & Pulse sequence \\
 \colrule
0 & $\textbf{A}-\text{Acq}$\\
1 & $\textbf{A}-\textbf{G}_1-\text{Acq}$\\
2 & $\textbf{A}-\textbf{G}_1-90_x-\textbf{G}_2-\text{Acq}$\\
3 & $\textbf{A}-\textbf{G}_1-90_x-\textbf{G}_2-54.74^{\circ}-\textbf{G}_3-\text{Acq}$\\
\end{tabular}
\end{ruledtabular}
\end{table}
The first set of experiments, listed in table \ref{experiment1},
comprises a stepwise progression through the twirl sequence,
equation \ref{2-qubit-3-axes-Gzonly}.  Each experiment uses the
same initial state, with $\tau=\units{69.3}{ms}$, a value chosen
to maximize the singlet fraction.  The three stages of the twirl
correspond to the application of three different crush gradients,
and can be characterized as follows: stage 0, do nothing; stage 1,
apply a single crush gradient; stage 2, apply the partial twirl
sequence \cite{anwar03}; stage 3, apply the full twirl sequence.
Two different measurements are performed to characterize the state
after each step: simple direct acquisition (which should show no
signals for a Werner singlet state) and acquisition after a
selective excitation pulse, which converts the singlet state into
NMR observable terms. The selective pulse was implemented using
the sequence
\begin{equation}\label{detection-D}
90_{45}-[\frac{1}{4\delta}]-90_{180},
\end{equation}
where $\delta=\units{915.8}{Hz}$ is the frequency separation
between the two resonances as before. This sequence is based on
jump-and-return sequences described in previous work
\cite{anwar03,jones99}, and when applied to a singlet state,
results in the observable NMR terms,
\begin{equation}\label{signal-singlet}
\frac{1}{2}(-2I_xS_z+2I_zS_x),
\end{equation}
corresponding to a pair of antiphase doublets with equal and
opposite intensities. The resulting spectra are shown in Figure
\ref{graphics1}. The left hand column of this figure shows spectra
from direct acquisition and the right hand column shows spectra
obtained with the excitation pulse followed by acquisition. The
four rows correspond to observation after 0, 1, 2 or all 3 stages
of the twirl sequence.

The initial state $\rho$ contains many different components, and
the observed spectra (top row) are complicated, reflecting this
fact. After the first stage of the twirl sequence (the first field
gradient) all components which are directly observable by NMR are
crushed (averaged to zero), and so no signal is visible in the
direct detection spectrum, but many other components remain,
indicated by the variety of signals seen after excitation.

The second stage of the twirl (the $90_x$ pulse and the second
gradient) removes most of these terms, producing a Bell diagonal
state with equal populations of $\ket{\Phi^+}$ and $\ket{\Phi^-}$.
As before there is no signal in the left hand spectrum, while the
spectrum on the right contains two antiphase doublets with clearly
different intensities.  This intensity difference arises from the
imbalance between the $\ket{\Psi^+}$ state and the two
$\ket{\Phi^\pm}$ states \cite{anwar03}.

After the third and final stage of the twirl (the $54.74_x$ pulse
and the last gradient) these two antiphase doublets have equal and
opposite intensities, characteristic of a Werner singlet state.
The final intensity of these doublets, compared with a calibration
spectrum (not shown) acquired from the thermal state, is
consistent with the twirl preserving the fraction of the singlet
state as expected.

\begin{figure}
\begin{center}
\includegraphics[scale=1.0]{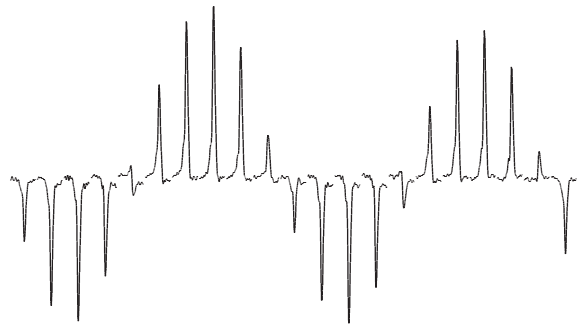}
\caption{Spectra from Werner singlet states obtained by twirling
impure states with varying singlet fractions; only one component of
one antiphase doublet is shown.  Successive spectra are obtained by
incrementing the variable delay in the preparation sequence,
Eq.~\ref{dirty-prepare}, in steps of
$1/(10\nu_I)=\text{\units{218}{$\mu$s}}$, which causes the fraction
of singlet in the initial state to be modulated, with a period of
ten steps.} \label{graphics2}
\end{center}
\end{figure}

This last point is explored in more detail in the second experiment,
in which the full twirl is applied to a range of initial states with
different amounts of the singlet state. The size of the singlet
component for our family of states is given in Equation
\ref{singlet-fraction-dirty}, and shows two kinds of sinusoidal
modulation with the variable delay $\tau$: a fast variation, arising
from the offset frequency $\nu_I$, on top of a slow variation due to
the coupling $J$ between the two qubits. If we choose $\tau$ to be
close to $1/(2J)$, we are near the maximum of the slow $J$
modulation, and the effect of varying $\tau$ is dominated by a
sinusoidal variation arising from $\nu_I$.

In this way we can produce a range of density matrices $\rho$ with
varying amounts of singlet, together with other terms, each of
which can be twirled to produce a Werner state.  This state can
then be observed using a selective excitation pulse prior to
acquisition.  In each case the expected spectrum is a pair of
antiphase doublets, with the intensity of the signal showing a
sinusoidal modulation at the frequency $\nu_I$.   The results of
this experiment are shown in Fig.~\ref{graphics2}, which depicts
the intensity variation in the righthand component of the lefthand
doublet as $\tau$ is varied around a value of
$1/(2J)=\units{69.3}{ms}$, with an increment between successive
spectra of $1/(10\nu_I)=\text{\units{218}{$\mu$s}}$; equivalent
effects can be seen for the other three components of the NMR
signal.  As expected a sinusoidal modulation of the signal is
clearly seen, and the observed modulation period of ten spectra is
exactly as expected.  The small out of phase signals observable
near the zero-crossings of the sine wave can be ascribed to the
effects of spin--lattice relaxation during the gradient pulses.

\section{Conclusions}
We have described a variety of strategies for the practical
implementation of twirl sequences on conventional and ensemble
quantum computers, and have demonstrated an ensemble
implementation on an NMR quantum computer.  With a conventional
quantum computer the implementation requiring the smallest number
of different bilateral operations is the set of 12 rotations
previously described \cite{bennett96b}, but our 18 step and 27
step averaging procedures require a smaller number of elementary
operations and may be simpler to implement in practice.  With an
ensemble quantum computer, such as an NMR device, it can be
simpler to replace the discrete averaging procedure by continuous
averaging, exploiting the ensemble nature of the system.  We have
developed a scheme, involving the application of three successive
crush gradients separated by RF pulses, which is well suited to
NMR quantum computers, and have demonstrated that its experimental
performance is consistent with theoretical expectations.

\begin{acknowledgments}
We thank the EPSRC and BBSRC for financial support.  MSA thanks the
Rhodes Trust for a Rhodes Scholarship.  HAC thanks MITACS for
financial support.
\end{acknowledgments}

\end{document}